\documentclass[usenatbib]{article}
\usepackage{times,graphicx,a4wide}

\begin{document}
	
\title{On the Possibility of Detecting Class A Stellar Engines Using Exoplanet Transit Curves}
\author{Duncan H. Forgan$^1$}
\maketitle

\noindent $^1$Scottish Universities Physics Alliance (SUPA), Institute for Astronomy, University of Edinburgh, Blackford Hill, Edinburgh, EH9 3HJ, UK \\

\noindent \textbf{Accepted for Publication in JBIS} \\
\noindent \textbf{Word Count: 6,587} \\

\noindent \textbf{Direct Correspondence to:} \\
D.H. Forgan \\ \\
\textbf{Email:} dhf@roe.ac.uk \\
\textbf{Post:} Dr Duncan Forgan \\
East Tower, Institute for Astronomy \\
University of Edinburgh \\
Blackford Hill \\
EH9 3HJ, UK

\newpage

\begin{abstract}

\noindent The Class A stellar engine (also known as a Shkadov thruster) is a spherical arc mirror, designed to use the impulse from a star's radiation pressure to generate a thrust force, perturbing the star's motion.  If this mirror obstructs part of the stellar disc during the transit of an exoplanet, then this may be detected by studying the shape of the transit light curve, presenting another potential means by which the action of extraterrestrial intelligence (ETI) can be discerned.  We model the light curves produced by exoplanets transiting a star which possesses a Shkadov thruster, and show how the parameters of the planet and the properties of the thruster can be disentangled. provided that radial velocity follow-up measurements are possible, and that other obscuring phenomena typical to exoplanet transit curves (such as the presence of starspots or intrinsic stellar noise) do not dominate.  These difficulties aside, we estimate the \emph{a priori} probability of detecting a Shkadov thruster during an exoplanet transit, which even given optimistic assumptions remains stubbornly low.  Despite this, many exoplanet transit surveys designed for radial velocity follow-up are on the horizon, so we argue that this remains a useful serendipitous SETI technique.  At worst, this technique will place an upper limit on the number of Class A stellar engines in the Solar neighbourhood; at best, this could help identify unusual transiting exoplanet systems as candidates for further investigation with other SETI methods.

\end{abstract}

Keywords: SETI, exoplanet, transits, megastructures

\section{Introduction}

\noindent For much of its sixty-year history, the Search for Extraterrestrial Intelligence (SETI) has relied heavily upon the detection of artificial radio signals emitted in the frequency range often referred to as the ``Water Hole'' or terrestrial microwave window \cite{Morrison1977}, a band of frequencies where the Earth's atmosphere is transparent and interference from Galactic and cosmic background radiation is minimal.  This band also happens to contain the 21 cm (or 1.42 GHz) spectral line emitted by neutral hydrogen atoms when the alignment of proton-electron spins "flip" from parallel to anti-parallel.  As hydrogen is the most abundant element in the Universe, it is often argued that extraterrestrial intelligences (ETIs) with radio technology will be scanning frequencies near to the 21cm line, and may even choose to emit  signals at frequencies related to it (e.g. 21$\pi$ cm or 21/$\pi$ cm).  This would suggest that Earth-like intelligent species could be detected by searching frequencies in the Water Hole for narrowband transmissions or broadband pulses \cite{Benford2010}.  

This motivation has inspired many radio SETI surveys carried out to date, placing constraints on the number of civilisations emitting in the surveyed bands in the Solar neighbourhood.  Most recently, a SETI survey of 86 stars in the Kepler field known to host transiting exoplanets estimates that less than 1\% of the targets host civilisations that emit in the 1-2 GHz band, at strengths comparable to those producible by humankind \cite{Siemion2013}.  Recent SETI searches using Very Long Baseline Interferometry (VLBI, \cite{Rampadarath2012}) show that the next generation of telescope arrays such as the Square Kilometre Array (SKA) will be able to survey the local neighbourhood to even higher fidelity, while efficiently discriminating potential signals from human radio frequency interference (see also \cite{Penny04}).

Equally, SETI scientists have recognised that searching in a limited region of the electromagnetic spectrum may be counter-productive to the overall goal of detection.  Even at the beginning of radio SETI, some argued for searches in the optical \cite{Schwartz1961}, noting that technology akin to the then recently-invented laser would be a highly efficient means of interstellar communication, both in its potential to traverse much larger distances at the same energy cost, and the ability to encode more information per unit time due to their higher carrier frequency.  While extinction by interstellar dust could destroy or mask a signal, ETIs can strike a balance by reducing the carrier frequency and emitting in the near infrared \cite{Townes1979}.  Recent searches for laser pulses in the spectra of local main sequence stars (e.g. \cite{Reines2002,Howard2004}) are also yet to bear fruitful detections. 

While both radio and optical SETI have their orthogonal advantages and disadvantages, they share a weakness common to all transient astrophysics - repeatability of detection.  The few cases where SETI searches uncovered a potential signal (e.g. the ``Wow!'' signal of 1977, \cite{Kraus1979,Gray2002}), the candidate signal was not rediscovered.  In the most memorable case where an apparently artificial signal was detected repeatedly, it was soon shown to be a previously undiscovered natural phenomenon - the pulsar \cite{Hewish1968,Penny2013}.  As pulsing transmitters are generally more energy efficient, there is a strong element of serendipity in detecting intentional transmissions, and as such surveys must be carefully designed \cite{Benford2010a, Messerschmitt2013}.  While it is possible that ETIs may deliberately repeat their signals to make their acquisition easier, they are under no compulsion to do so.

If the goal is to detect unintentional transmissions, then this obstacle can be avoided, as these transmissions are expected to be produced as part of the typical activities of an extraterrestrial civilisation.  Unfortunately, unintentional transmissions by their very nature are not designed to travel large interstellar distances - terrestrial radio leakage could be detected by current and future instrumentation at no more than a few hundred parsecs \cite{Loeb2007}, and their content is likely to be less informative than that transmitted by a dedicated beacon.  Indeed, in the case of radio signal leakage, it may only be possible to detect civilisations in their earliest uses of radio communication, as they may wish to subsequently reduce signal leakage as a cost-cutting measure \cite{SKA}, or to shield themselves from dangerous eavesdroppers \cite{Shostak2012,deVladar2012}.

However, just as a planetary biosphere displays atmospheric biomarkers that can be used in some cases to infer the presence of life \cite{Lovelock1975,Virgo2012}, planetary systems inhabited by intelligent species are likely to present \emph{noomarkers} (or sometimes ``technomarkers'') which are difficult to hide.  Using the classification system for civilisations proposed by \cite{Kardashev1964}, these noomarkers are produced strongly by Type II civilisations, which manipulate energy resources equivalent to typical stellar luminosities, rather than the Type I civilisations more reminiscent of humankind, whose access to energy resources is limited to the much smaller planetary scale.  Having said this, we should note that noomarkers produced by asteroid mining \cite{asteroid_mining}, may be present in the Solar System, albeit weakly, before human civilisation reaches the Type II classification, if recently founded commercial ventures to mine asteroids are successful (see e.g. http://deepspaceindustries.com/ and http://www.planetaryresources.com/).

Megastructures on the scale of celestial objects constitute the most studied type of noomarker, the classic example being the Dyson sphere \cite{Dyson}.  Indeed, SETI searches for this type of artifact \cite{Slysh1985,Timofeev2000,Carrigan2009} are often referred to as ``Dysonian SETI'' in deference to its most well-known advocate \cite{Cirkovic_macro}.  Megastructures or macro-engineering projects are motivated by a Malthusian view of the future development of humanity.  Rising demand from an increasing population eventually encounters the paucity of supply, as terrestrial resources are exhausted.  When this happens, the high costs of macro-engineering projects can soon be outweighed by the enormous benefits they can bring, not to mention the predicted costs of \emph{not} carrying out the project \cite{Hickman_economy}.  This is hardly an ironclad proof that ETIs will always build megastructures, and one can be easily led astray when speculating on the motives of advanced technological civilisations, but at the very least it suggests that there are some evolutionary trajectories for technological civilisations that include the construction of Dysonian-type artifacts, especially if a civilisation adopts an optimisation-driven rather than expansion-driven developmental strategy \cite{empire}.

There have been many suggestions of other noomarkers that are amenable to SETI searches, such as pollution in planetary \cite{Campbell2006} and stellar atmospheres \cite{Whitmire1980}, or deliberate ``salting'' of the host star with unusual isotopes (e.g. \cite{Shklovsky1966,Carrigan2010}).  These suggestions are motivated by the fact that there is a great deal of study of stars and planets for their own sake. 

From the first detection of an extrasolar planet (or exoplanet) around a main sequence star almost two decades ago \cite{Mayor1995}, there has been a drive to discover and characterise a large number of exoplanets, as the properties of the exoplanet population allows astronomers to understand the physical processes at work in planet formation.  While exoplanets can be detected in a variety of ways (see \cite{Wright2012} for a general review), the majority of the known exoplanet population has been detected using one or (or both) of the following methods.  The \emph{radial velocity} method (cf \cite{Lovis2011}) studies the star's spectral lines for evidence of changes in velocity along the line of sight to the observer.  As stars and planets orbit a common centre of mass, the evolution of the star's radial velocity encodes information regarding the planet's orbital period and semi-major axis, as well as the minimum mass of the planet, $M_p \sin i$, where $M_p$ is the true mass and $i$ is the inclination of the orbit (with $i=90^{\circ}$ equivalent to the star, planet and observer existing in the same plane).  The transit method (cf \cite{Winn2011}) requires the observer to study the star's flux as a function of time.  If the planet passes between the star and the observer, this produces a dimming of the stellar flux proportional to the size of the silhouette cast by the planet on the stellar disc.  The transit method constrains the planet's physical radius $r_p$ relative to the stellar radius $R_*$.  The combination of both methods allows the inclination to be measured, and the true mass $M_p$ to be determined.    

While other detection methods exist, such as gravitational microlensing (e.g. \cite{Gaudi2011}) or direct imaging (e.g. \cite{Wright2012}), radial velocity and transit studies have not only revealed the largest quantity of exoplanets to date, but also the most detailed data on individual exoplanet systems, when both methods are used in tandem.

Exoplanet detection benefits SETI in several ways; firstly, it provides invaluable data on the frequency of potentially habitable terrestrial planets, constraining one of the seven (or more) terms of the master equation of SETI, the Drake Equation, which is cast in various guises \cite{Bracewell1976,Drake_eqn, Maccone2011,Glade2012}.  Secondly, exoplanet studies can in principle identify objects which are highly likely to not be merely habitable, but inhabited.  For example, exoplanet transit spectroscopy, which uses multi-wavelength observations of transits to study the planet's atmospheric absorption as a function of wavelength, is likely to prove effective as a probe for the presence of biomarkers (\cite{Hedelt2013} and references within). Thirdly, it provides an extensive, rich dataset of astrophysical phenomena on planetary system scales, which can be studied in a serendipitous or ``piggyback" fashion for evidence of artificial signals or noomarkers.  

In particular, the transit detection method allows the exciting possibility of indirectly detecting megastructures through their obscuration of the parent star.  \cite{Arnold2005} illustrates how transit curves could contain a deliberate signal placed by ETIs in the form of large geometric structures, which have transit signatures distinct to that of an occulting spherical disc.  Such structures could be placed in an orbit where observing civilisations would be able to easily detect an exoplanet transit (cf \cite{Conn2008}), with the lifetime of the resulting signal potentially very long indeed \cite{Arnold2013}. 

In this paper, we focus on a particular class of megastructure, known as the Class A stellar engine, or the Shkadov thruster, in honour of its first description by \cite{Shkadov1987}.  This object is essentially a mirror which reflects a fraction of the star's radiation pressure, causing a force asymmetry which exerts a thrust on the star.  This thrust could be used to move a civilisation's host star from its ``natural'' orbit if it posed some harm to the civilisation (e.g. a dangerous close approach to another star or dust cloud).  We study the potential transit signatures produced by an exoplanet in a system that contains such an engine, and show how to disentangle the planet's contribution to the transit from the engine's occultation of the star.  In section \ref{sec:engine} we describe the physics of the Shkadov Thruster; in section \ref{sec:transit} we recap the structure of a transit curve in the absence of a stellar engine; in section \ref{sec:transit_engine} we show how the transit curve is modified in the presence of a Shkadov thruster, and how to analyse these curves in section \ref{sec:deconstruct}; in section \ref{sec:discussion} we discuss the implications of this work and we summarise our conclusions in section \ref{sec:conclusion}.

%

\section{Stellar Engines and the Shkadov thruster\label{sec:engine}}

\noindent In short, a stellar engine is any device or structure which extracts significant resources from a star in order to generate work.  Typically, the resource in question is the radiation field produced by the star, although some stellar engines use the mass of the star as propellant, e.g. the  stellar rockets or ``star lifters'' as proposed by \cite{Criswell1986} and \cite{Fogg1989}.  

Strictly speaking, the most well known theoretical megastructure, the Dyson sphere \cite{Dyson}, is not a stellar engine, as it does not specifically generate work.  The Dyson sphere is a spherical shell that englobes the star, with a radius and thickness such that the radiation pressure force from the star and the gravitational force on the sphere are in balance, producing a static satellite (or \emph{statite}).  While usually envisaged as a single solid structure, it can be considered as a very large collection of smaller bodies, arranged in a spherical geometry.  This is sometimes referred to instead as a Dyson swarm, and is used only to collect energy.  

The Dyson sphere/swarm's function is therefore to either a) collect stellar energy, or b) provide living space, or both.  The interior of a single solid Dyson sphere, designed as an Earth-type habitat, will provide a significantly larger amount of living space compared to a planetary body, as the radius of a Dyson Sphere will be very large relative to terrestrial planet radii.

\begin{figure}
\begin{center}
\includegraphics[scale = 0.3]{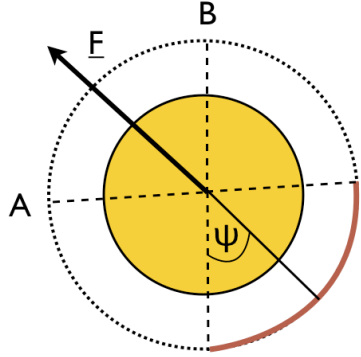} 
\caption{Diagram of a Class A Stellar Engine, or Shkadov thruster.  The star is viewed from the pole - the thruster is a spherical arc mirror (solid line), spanning a sector of total angular extent $2\psi$.  This produces an imbalance in the radiation pressure force produced by the star, resulting in a net thrust in the direction of the arrow.  \label{fig:shkadov_thruster}} 
\end{center}
\end{figure}

While the Dyson sphere is not a stellar engine, most stellar engine designs share similarities with the Dysonian model - in particular, they share the Dyson sphere's hunger for construction material.  A full Dyson sphere would require the systematic destruction of several planetary bodies to provide raw materials \cite{Dyson,Badescu2000}.  \cite{Badescu2000} present a classification system for stellar engines: we will focus exclusively on their Class A stellar engine, or Shkadov thruster (see Figure \ref{fig:shkadov_thruster}).  

\noindent A spherical arc mirror (of semi-angle $\psi$) is placed such that the radiation pressure force generated by the stellar radiation field on its surface is matched by the gravitational force of the star on the mirror.  Radiation impinging on the mirror is reflected back towards the star, preventing it from escaping.  This force imbalance produces a thrust of magnitude \cite{Badescu2000}:

\begin{equation}
F = \frac{L}{2c}\left(1-\cos \psi\right),
\end{equation}

\noindent where $L$ is the stellar luminosity.  This result is similar to that produced by \cite{Shkadov1987}, although their calculation assumes the temperature of the star remains constant.  In reality, the reflected radiation will alter the thermal equilibrium of the star, raising its temperature and producing the above dependence on semi-angle.  Increasing $\psi$ increases the thrust, as expected, with the maximum thrust being generated at $\psi=\pi$ radians.  However, if the thruster is part of a multi-component megastructure that includes concentric Dyson spheres forming a thermal engine, having a large $\psi$ can result in the concentric spheres possessing poorer thermal efficiency \cite{Badescu2000}.

\section{Exoplanet Transit Curves \label{sec:transit}}

\noindent We briefly recap the typical shape of an exoplanet transit curve here (for a review see e.g. \cite{Winn2011})  Rather than attempting to simulate observations by pixelating the stellar disc and carrying out parameter fitting, as was done by \cite{Arnold2005}, we instead construct theoretical curves in the following fashion.

We work in $(x,y)$ coordinates, where the coordinates represent the projection of the planetary system onto a two-dimensional ``observer'' plane.  A single star is placed at the origin $(0,0)$, possessing a stellar disc with total area $A_*=1$, i.e its radius is 

\begin{equation}
R_*  = \sqrt{\frac{A_*}{\pi}} = \sqrt{\frac{1}{\pi}}
\end{equation}

in arbitrary units.  A planet (with a disc area $A_p=0.01$, and radius $r_p = \sqrt{0.01/\pi}$) is positioned initially at $(x_i,b)$, where $b$ is the impact parameter.  We assume that the projection of the planet's orbit on the plane is a straight line rather than an arc, and move the planet in the x-axis only, across the stellar disc until it reaches $(x_f,b)$.  As transit curves typically normalise the received flux by the steady value obtained during a non-transiting period of the planetary orbit, we calculate the flux using

\begin{equation}
F = 1.0 - \frac{A_{int}}{A_*}
\end{equation}

\noindent Where $A_{int}$ is the area inside the intersection of the stellar and planetary discs, and $A_{int} = [0, A_p]$.  By construction, the minimum flux is 0.99, and we neglect thermal emission from the planet.  $A_{int}$ is analytically calculable provided the positions of the star and planet are known, by solving a quartic equation for the intersection points of the two discs, and calculating the circular segment area subtended inside each disc between the intersection points.

The transit curve can be deconstructed into several components, delineated by four events where the planetary and stellar discs possess one intersection point only, i.e. the discs are in tangential contact:

\begin{enumerate}
\item Beginning of Transit Ingress ($\tau_{I}$) - where the star is as yet unaffected by transit.
\item End of Transit Ingress ($\tau_{II}$) - the planet is now fully within the stellar disc.  In the absence of stellar limb-darkening, the light curve would reach its minimum value here.
\item Beginning of Transit Egress ($\tau_{III}$) - the planet is now moving outside the stellar disc, and the received stellar flux increases, and finally
\item End of Transit Egress ($\tau_{IV}$) - the planet no longer obscures the stellar disc, and the transit is complete.
\end{enumerate}

\begin{figure*}
\begin{center}$
\begin{array}{cc}
\includegraphics[scale = 0.4]{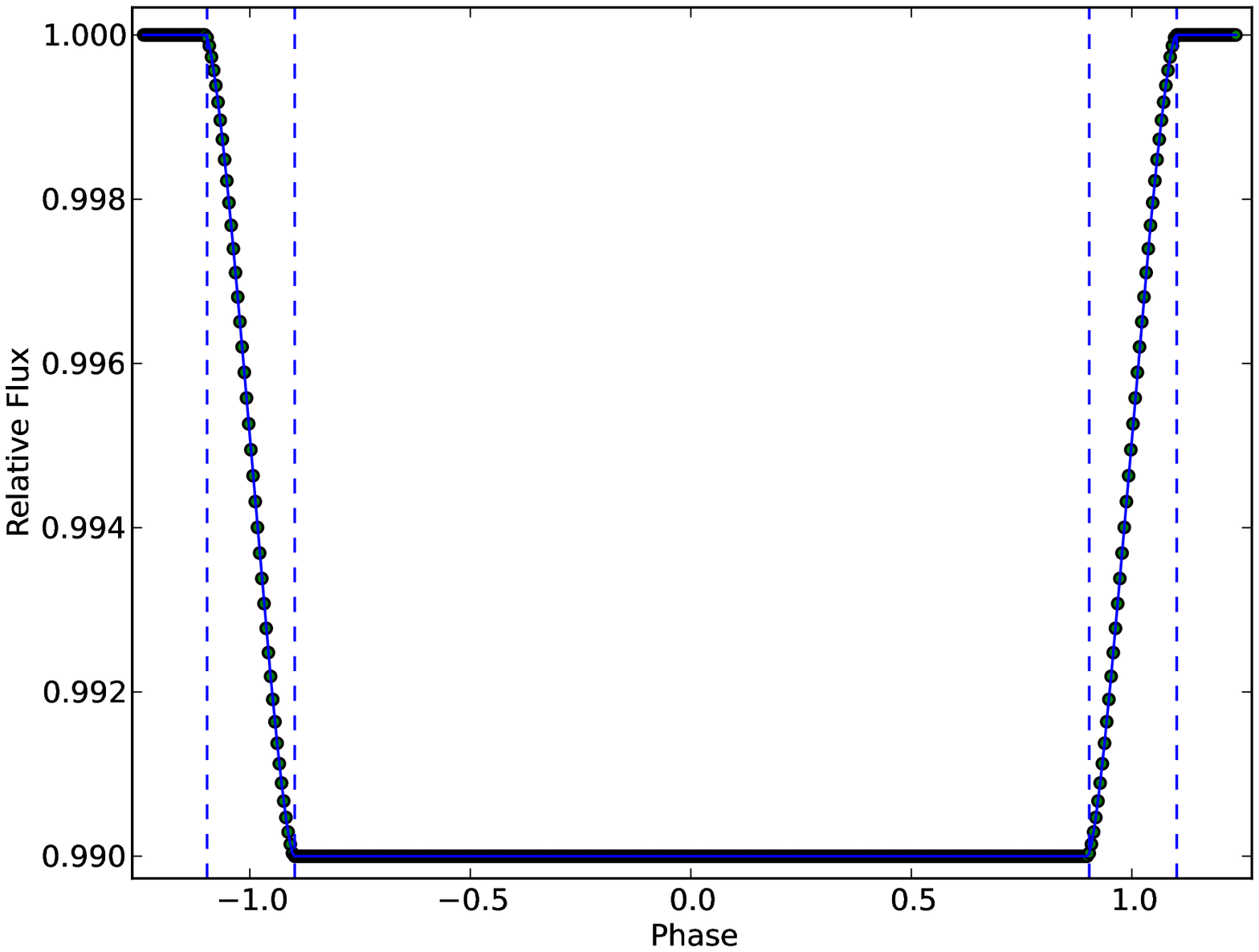} &
\includegraphics[scale=0.4]{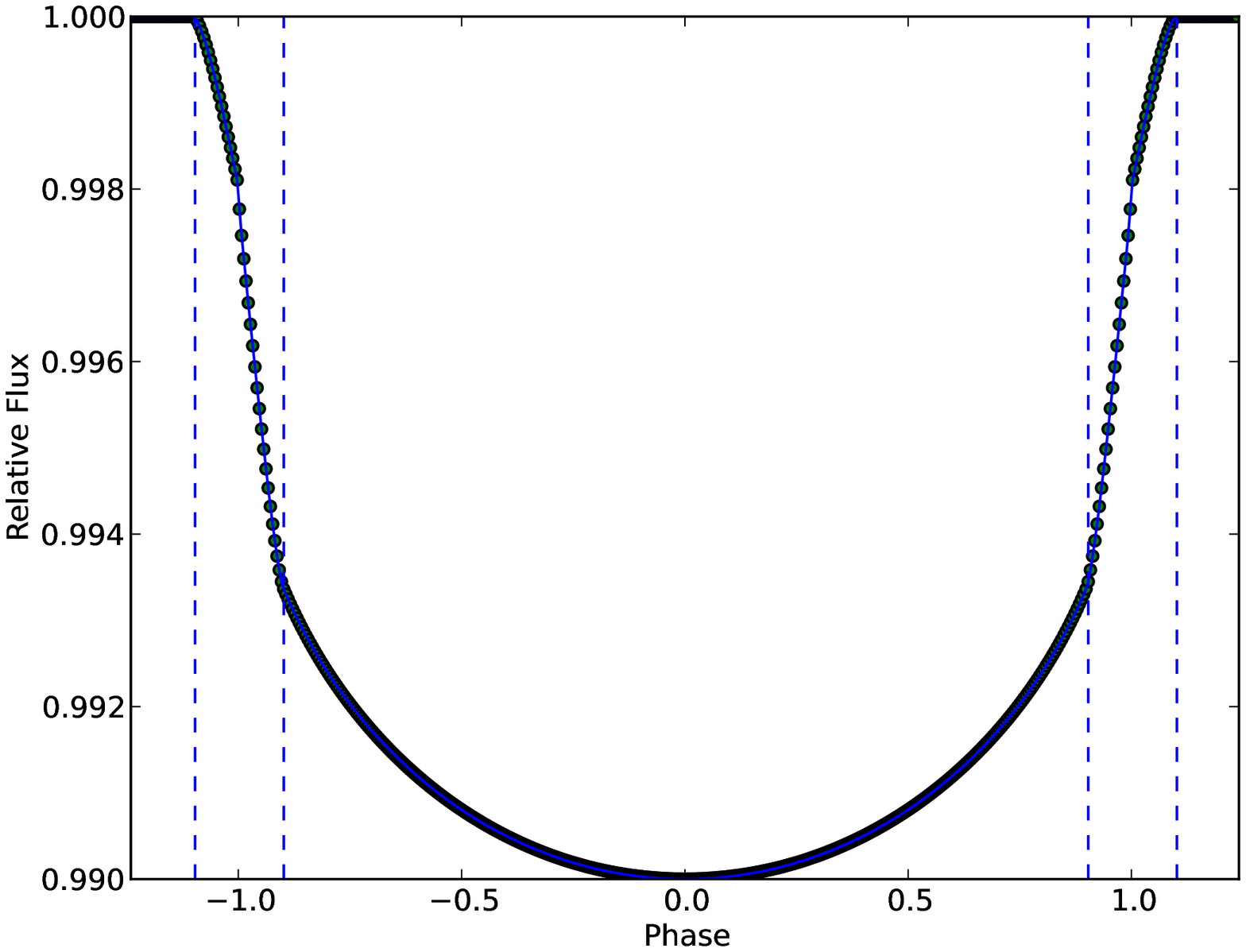} \\
\end{array}$
\caption{Examples of the transit curves produced for a planet without a stellar engine present (with impact parameter $b=0$).  The left panel shows the curve in the absence of stellar limb-darkening, and the right panel shows the curve with limb-darkening.  The dotted lines in each plot represent (from left to right): beginning of ingress ($\tau_{I}$), end of ingress ($\tau_{II}$), beginning of egress ($\tau_{III}$) and end of egress ($\tau_{IV}$). \label{fig:transit_b0}} 
\end{center}
\end{figure*}

\noindent Figure \ref{fig:transit_b0} shows the transit curves produced both in the absence and presence of stellar limb-darkening.  We do not model the curves as a function of time, but rather as a function of planetary position.  We use a dimensionless phase parameter $\phi$ for the x-axis in this and all subsequent curves, where $\phi=x_p/R_*$, and $x_p$ is the x-coordinate of the planet's centre.  In these units, the centre of the planet touches the stellar perimeter at $\phi=-1,1$.

To simulate limb-darkening, we modify $A_{int}$ using

\begin{equation}
A_{int} \rightarrow A_{int}\left(1 - u(1- \mu) \right) \label{eq:limbdark}
\end{equation}

\noindent where 

\begin{equation}
\mu = \sqrt{\frac{R^2_* - d^2}{R^2_*}},
\end{equation}

\noindent the distance between the centres of the stellar and planetary discs is denoted by $d$, and we select the limb-darkening parameter $u=0.6$ (where $u$ is typically a function of the observing wavelength).  Equation \ref{eq:limbdark} is one of many different expressions used for limb-darkening which are derived from stellar-atmosphere models (see \cite{Claret2004} for a list), but this work is not particularly sensitive to which parametrisation that we adopt, but merely requires that we adopt one.  Increasing the dimensionless impact parameter 

\begin{equation}
b = \frac{a \cos i}{R_*}
\end{equation}

\noindent can partially mimic the curve shape in the ingress-egress sections that is also produced by the limb-darkening.  However, both parameters can usually be disentangled with an appropriate best-fit modelling procedure (see e.g. \cite{Winn2011}).  If the oblateness of the planet is non-zero, or the planet possesses rings, then this can be detected provided the photometric precision is better than $\sim 10^{-4}$ \cite{Seager2002,Barnes2003}.  


\noindent Let us define the transit depth (i.e. the fractional difference in flux before and during the transit) as $\delta$ and the ratio of the planet to stellar radius $r_p/R_* \equiv k$.  As such, if both the stellar and planetary discs are circular,

\begin{equation}
\delta = k^2.
\end{equation}

\noindent While in this work we model the transit curve as a function of planetary position, not time, it will be useful to define the equatorial crossing timescale (for a circular orbit):

\begin{equation}
\tau_{eq} = \frac{R_* P_{pl}}{2\pi a_p} 
\end{equation}

\noindent Where $P_{pl}$ is the planet's orbital period, and $a_p$ is the planet's semi-major axis.  The total transit time is 

\begin{equation}
T_{tot} = \frac{P}{\pi} \sin^{-1} \left(\frac{R_*}{a} \sqrt{1+k^2-b^2}\right)
\end{equation}

\noindent We can therefore convert our dimensionless phase into real time by using

\begin{equation}
t = \frac{\phi T_{tot}}{2}, 
\end{equation}

\noindent where $t=0$ represents the midpoint of the transit.  If we assume that $r_p << R_* << a$, and $b<< 1-k$ (i.e. we exclude transits that come close to grazing the upper or lower limb of the star), the ingress timescale is:

\begin{equation}
\tau_{ing} \equiv \tau_{II}-\tau_{I} =  \tau_{eq} \left(\sqrt{(\left(1+ k\right)^2 -b^2} +\sqrt{(\left(1- k\right)^2 -b^2}\right) \sim 2 \tau_{eq} \frac{k}{\sqrt{1-b^2}} 
\end{equation}

\noindent Armed with an appropriate model to fit stellar limb darkening and remove sources of false positives such as starspots, the transit curve, specifically the parameters $\left(\delta, T_{tot}, \tau_{ing}\right)$, is sufficient to characterise $k$ and $b$.  Stellar atmosphere modelling, or in some cases asteroseismology \cite{Stello2009} can then be used to define $R_*$ to obtain $r_p = kR_*$.

\section{Transit Curves in the presence of a Class A Stellar Engine \label{sec:transit_engine}}
		
\begin{figure}
\begin{center}
\includegraphics[scale = 0.3]{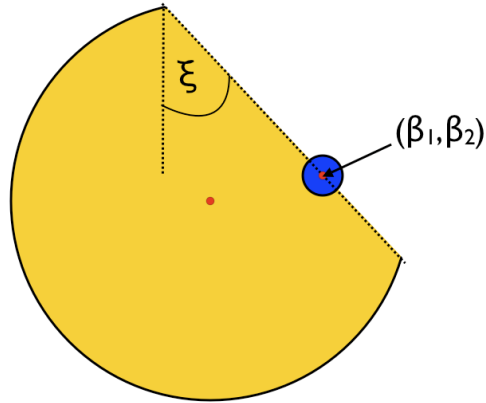} 
\caption{Diagram of a transit in the presence of a Class A Stellar Engine or Shkadov Thruster (which we will refer to throughout as a ``Shkadov transit'').  The Shkadov thruster subtends a chord at angle $\xi$ to the y-axis, and obscures an area of the stellar disc $\Sigma_s$, in this case to the right of the chord.  The chord passes through a point $(\beta_1,\beta_2)$ on the stellar disc (in the diagram, $\beta_2=b$, the planet's impact parameter).  As the thruster is a highly reflective mirror, there is no limb darkening associated with it.\label{fig:shkadov_transit_graphic}} 
\end{center}
\end{figure}

\noindent We model the presence of an obscuring Class A Stellar Engine as a circular segment drawn on the stellar disc at an angle $\xi$ to the y-axis, passing through some point on the stellar disc $(\beta_1,\beta_2)$ (see Figure \ref{fig:shkadov_transit_graphic}).  In practice, we will measure these values normalised to the stellar radius in the same manner as the transit's impact parameter, i.e. $\beta_1,\beta_2= [0,1]$.

The engine completely obscures the star to the right hand side of the chord, and as a result when the planetary disc passes into the region where the star is obscured, the transit curve will be altered, particularly its egress features.  Note that throughout this work we position the engine in the positive x-axis region of the star without affecting the generality of the result - engines in the negative x region would simply produce signatures at the curve's ingress point rather than the egress point.  

\begin{figure}
\begin{center}
\includegraphics[scale = 0.4]{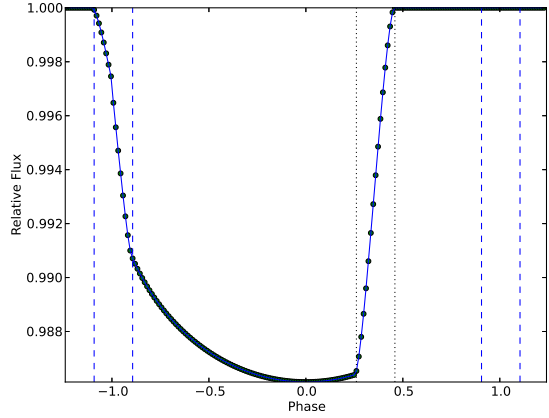} 
\caption{The light curve associated with a Shkadov Transit.  The dashed lines represent the expected ingress/egress points for a transit without a Shkadov thruster present, and the dotted lines represent the beginning and end points of the planet's ingress into the region obscured by the Shkadov thruster, which we refer to as the ``Shkadov egress''. \label{fig:shkadov_transit_example}} 
\end{center}
\end{figure}

We define the obscured area of the stellar disc as $\Sigma_s$.  As the engine is a highly reflective mirror, there is no limb-darkening associated with the obscured side of the star.  The depth of the transit curve will no longer be equivalent to the ratio of the planetary and stellar disc areas, but will also depend on $\Sigma_s$:

\begin{equation}
F = 1.0 - \frac{A_{int}-A_{int,s}}{A_* - \Sigma_s}, \label{eq:F_shkadov}
\end{equation}

where $A_{int,s}$ is the area inside the intersection of the planetary disc and the Shkadov thruster.  When the planet passes into the region where the stellar disc is obscured, the transit curve will show the planet beginning ``egress'' earlier than in the case where the engine is absent, introducing an asymmetry (see Figure \ref{fig:shkadov_transit_example}).  We will hereafter refer to transits of this type as ``Shkadov transits", and we will refer to the apparently early egress of the curve as the ``Shkadov egress''.  Although it is strictly also correct to refer to it as the Shkadov ingress, it is perhaps more sensible to allow each curve to have an ingress and egress component rather than two of the same.  As the ingress section of the curve is limb-darkened and the Shkadov egress is not, the ingress and egress will progress with different measured gradients. 

While these curves of this type are difficult to produce naturally, and as such are \emph{prima facie} evidence for macro-engineering occuring in this particular star system, what else can be divined from the curve? Specifically, can we determine the properties of both the stellar engine as well as the true exoplanet parameters? After all, the curve's symmetry is lost.  We cannot measure $T_{tot}$; the depth of the Shkadov transit curve, $\tilde{\delta}$, is now degenerate in $r_p$ and $\Sigma_s$: 

\begin{equation}
\tilde{\delta} = \frac{k^2}{1-\Sigma_s/\left(\pi R^2_*\right)},
\end{equation}

\noindent In other words, what can we now measure to recover $(k,b)$ and the properties of the thruster?  

\begin{figure*}
\begin{center}$
\begin{array}{cc}
\includegraphics[scale = 0.4]{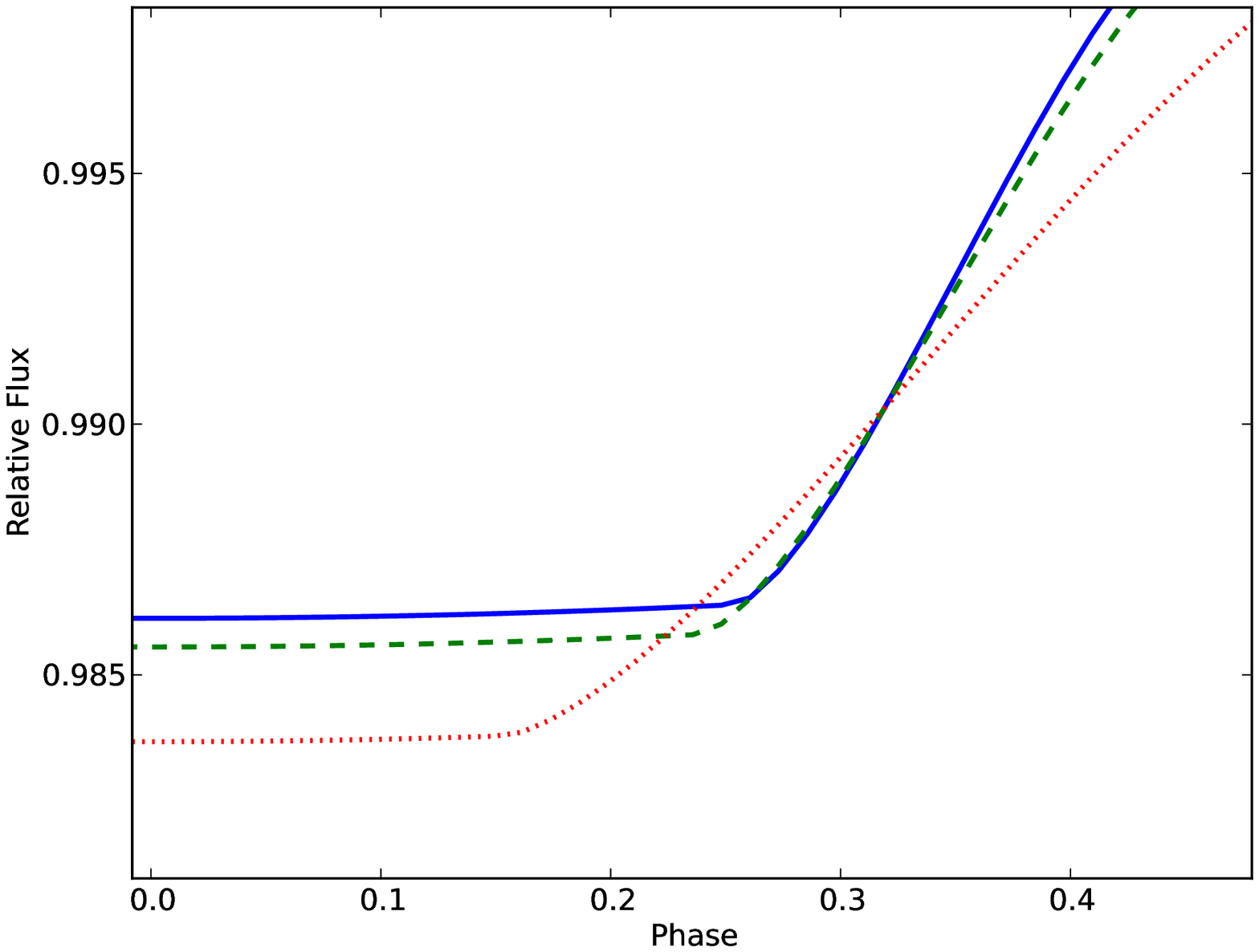} &
\includegraphics[scale=0.4]{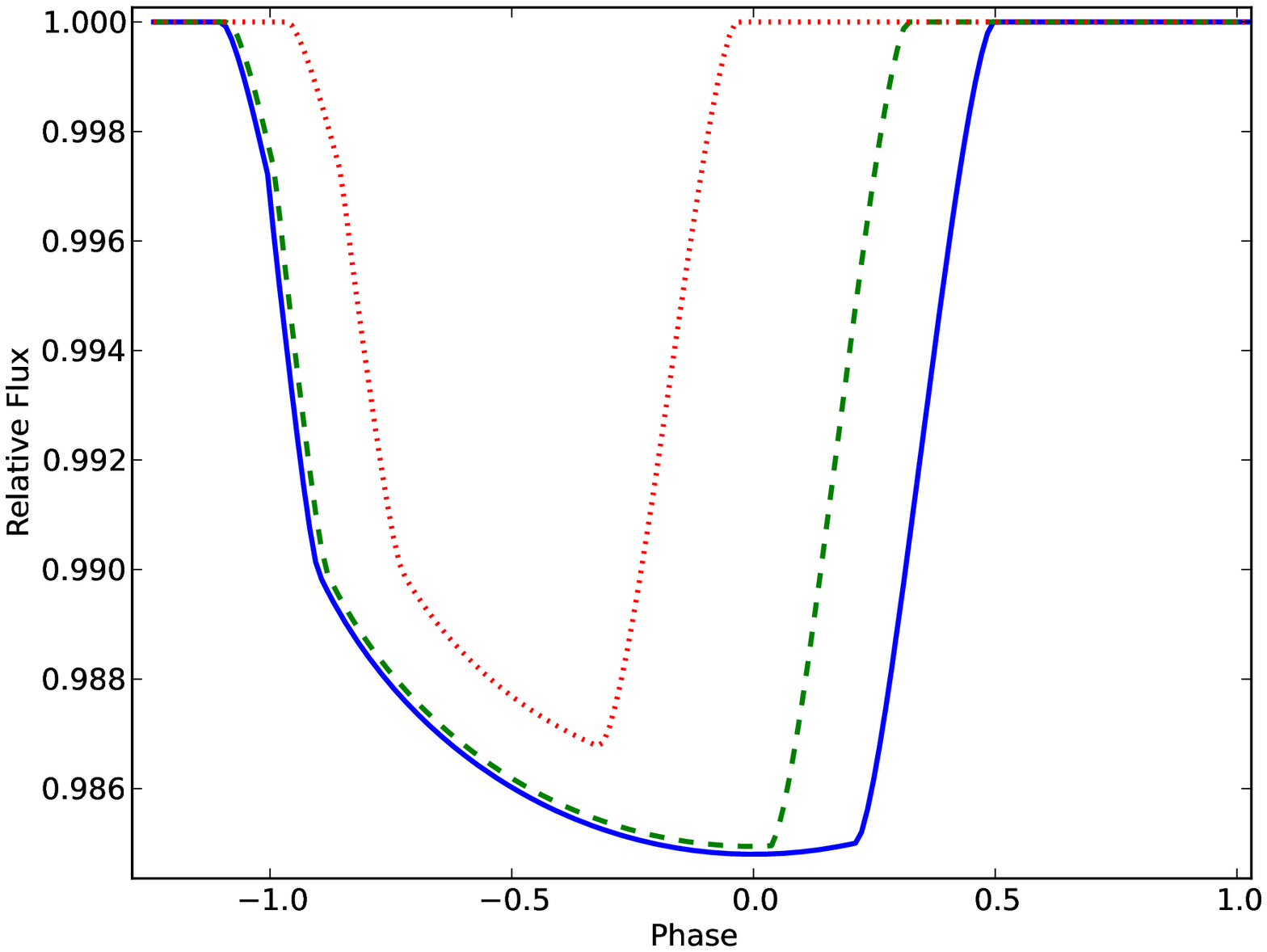} \\
\end{array}$
\caption{Left: The egress section of the light curve during Shkadov transits for a planet with impact parameter $b=0$, as the angle between the engine and the y-axis, $\xi$, is increased.  The dotted curve (with the deepest transit depth) has $\xi=0^{\circ}$; the dashed curve with moderate depth has $\xi=30^{\circ}$, and the shallow solid curve is $\xi=60^{\circ}$.  Right: The dependence of the Shkadov transit light curve as the planetary impact parameter, $b$, is increased, for an engine of fixed $\xi=45^{\circ}$ and $(\beta_1,\beta_2) = (0.2,0)$.  The solid curve (with the longest transit duration) has $b=0$; the dashed curve with moderate duration has $b=0.1$, and the dotted curve with shortest duration has $b=0.3$. \label{fig:shkadov_xi_b_curves}} 
\end{center}
\end{figure*}

We can see the beginning of the solution to this problem by holding $(\beta_1,\beta_2)$ fixed and varying $\xi$ (left hand panel of Figure \ref{fig:shkadov_xi_b_curves}).  The midpoint of ingress into the Shkadov thruster is fixed at $\beta_1$ (note all curves intersect at a phase equal to $\beta_1$).  Increasing $\xi$ while holding $(\beta_1,\beta_2)$ fixed reduces $\Sigma_s$, and the duration of the Shkadov egress also increases.

\noindent If we now consider the engine having fixed values of $\xi=45^{\circ}$ and $(\beta_1,\beta_2)$, and instead vary the planet's impact parameter $b$, then we can see (right hand panel of Figure \ref{fig:shkadov_xi_b_curves}) how the presence of the engine will cut the transit short depending on its value of $b$.  Planets with higher $b$ will encounter the obscured area earlier, and hence egress begins earlier.  As we model the silhouette of the engine's edge as a straight line, all the curves in the right hand panel of Figure \ref{fig:shkadov_xi_b_curves} show the same duration of egress.


\section{Deconstruction of A Shkadov Transit Curve \label{sec:deconstruct}}

Provided that:

\begin{description}
\item[i] ingress of the exoplanet into the stellar disc is not also obscured,
\item[ii] radial velocity follow-up measurements are possible, and 
\item[iii] the radius of the star $R_*$ is still measurable (through e.g. asteroseismology measurements),
\end{description}

\noindent this should be sufficient to estimate the transiting timescale $\tau_{eq}$, the ingress timescale $\tau_{ing}$ and hence the degenerate $k/\sqrt{1-b^2}$.   

If the orbital period $P_{pl}$ and the semi-major axis $a_p$ are well-constrained as a result of the combined radial velocity/transit measurements, then the transit curve should be sufficiently informative to break the degeneracies linking $(r_p, \xi, \beta_1,\beta_2)$.  The measurements to be made are: the Shkadov egress timescale $\tilde{\tau}_{eg}$, the Shkadov transit depth $\tilde{\delta}$, and the midpoint of Shkadov egress $\phi_{eg} = \beta_1$.  At the midpoint of Shkadov egress, $\beta_2=b$, reducing the number of parameters to estimate by one.

\subsection{The Shkadov Egress Timescale}

\begin{figure}
\begin{center}
\includegraphics[scale = 0.3]{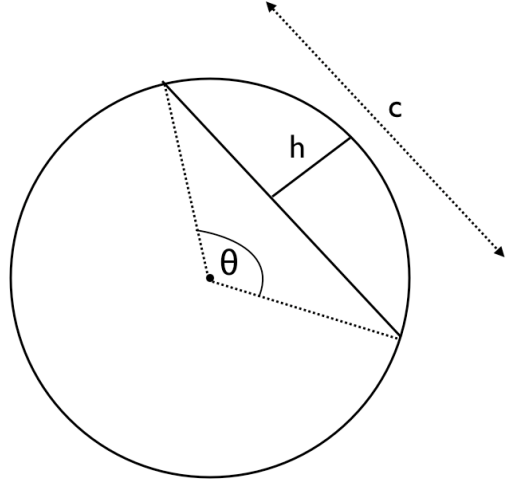} 
\caption{Diagram of a circular segment, with central angle $\theta$, height $h$ and chord length $c$. \label{fig:shkadov_segment}} 
\end{center}
\end{figure}

\noindent As can be seen from Figure \ref{fig:shkadov_xi_b_curves}, the time taken for the planet to complete Shkadov egress is independent of $b$, and depends only on $\xi$.  If we define the Shkadov thruster as a circular segment of central angle $\theta$, and height $h$ (see Figure \ref{fig:shkadov_segment}), then we can analytically calculate how the area of the planetary disc obscured by the thruster changes as the planet sweeps across the chord.

\noindent The height $h$ of a circular segment (inside a circle of radius $R$) and the central angle that spans the segment, $\theta$ are related by:

\begin{equation}
h = R \left(1 - \cos \left(\frac{\theta}{2}\right)\right). \label{eq:h_theta}
\end{equation}

\noindent The length of the chord defining the segment is

\begin{equation}
c = 2R\sin \left(\frac{\theta}{2}\right),
\end{equation}

\noindent and the area of the segment is

\begin{equation}
A(\theta) = \frac{R^2}{2} \left(\theta - \sin \theta\right). \label{eq:A_theta}
\end{equation}

\noindent As the planetary disc sweeps across the thruster chord, the thruster obscures a circular segment of height $h_p(t)$ and angle $\theta_p(t)$.  If the thruster has $\xi=0^{\circ}$, then $h_p$ is parallel to the $x$ axis and

\begin{equation}
\frac{d h_p}{dt} = \dot{h}_p = \dot{x}.
\end{equation}

\noindent It can be shown that if $\xi\neq 0^\circ$, then

\begin{equation}
\dot{h}_p =\dot{x}  \cos \xi .
\end{equation}

\noindent Differentiating equation (\ref{eq:h_theta}) gives

\begin{equation}
\dot{h}_p = \left(\frac{R_*}{2} \sin \theta_p\right) \dot{\theta}_p = \frac{c}{4} \dot{\theta}_p \label{eq:hdot_thetadot}
\end{equation}

\noindent Applying equation (\ref{eq:A_theta}) to the planetary disc, and taking the first derivative gives

\begin{equation} 
\dot{A} = \frac{r^2_p}{2} \left(1 - \cos \theta_p \right)\dot{\theta}_p.
\end{equation}

\noindent To solve this, we must solve for $\dot{\theta}_p$.  From equation (\ref{eq:hdot_thetadot}):

\begin{equation}
\dot{\theta}_p -  \frac{4 \dot{h}_p}{c} = \dot{\theta}_p - \frac{4 \dot{h}_p}{2 r_p \sin \left(\frac{\dot{\theta}_p}{2}\right)} = 0.
\end{equation}

\noindent Differential equations of the form

\begin{equation}
\dot{y} - \frac{A}{\sin\left(\frac{y}{2}\right)} =0.
\end{equation}

\noindent Have the solution

\begin{equation}
y(t) = \pm 2 \cos^{-1}\left(\frac{1}{2} \left(-At -C\right)\right),
\end{equation}

\noindent where $C$ is a constant of integration.  If we impose the initial condition $y(0)=0$, then $C=-2$.  Substituting for $A=2\dot{h}_p/r_p$, and assuming $\dot{x}$ is constant:

\begin{equation}
\cos \frac{\theta_p(t)}{2} = 1 - \frac{\dot{h}_p}{r_p} t =  1 - \frac{\cos \xi \dot{x}}{r_p} t.
\end{equation}

\noindent If we assume that 

\begin{equation}
\dot{x}=\frac{2\pi a_p}{P_{pl}},
\end{equation} 

\noindent then we can calculate the egress timescale as the time at which $\theta_p=2\pi$, or

\begin{equation}
\tilde{\tau}_{eg} = \frac{P}{\pi a_p}\frac{r_p}{\cos \xi}.
\end{equation}

\noindent Therefore, the egress timescale allows us to constrain the value of $r_p/\cos \xi$.

\subsection{Calculating $\Sigma_s$}

\noindent To compute $r_p$ from $\tilde{\delta}$, we must be able to compute $\Sigma_s$, which we can if we know $(\beta_1,\beta_2)$ and $\xi$.  If we define $\mathbf{i} = (i_1, i_2)$ and $\mathbf{f} = (f_1,f_2)$ as the points of intersection between the chord and the stellar disc, normalised to the stellar radius, then

\begin{equation}
i^2_1 +i^2_2 = f^2_1 +f^2_2 = 1
\end{equation}

\noindent The points of intersection must obey the vector equation

\begin{equation}
\mathbf{i} + c\mathbf{\hat{c}} = \mathbf{f}
\end{equation}

\noindent Where $\mathbf{\hat{c}} \equiv \left(\sin \xi, \cos \xi\right)$ is the unit vector with direction along the chord.

\noindent  As the point $\mathbf{\beta} = (\beta_1,\beta_2)$ rests on the chord, we can write 

\begin{equation}
\mathbf{i} + c_1\mathbf{\hat{c}} = \mathbf{\beta}
\end{equation}

\begin{equation}
\mathbf{\beta} + c_2\mathbf{\hat{c}} = \mathbf{f}
\end{equation}

\noindent where $c_1 + c_2 = c$.  If $\xi$ is known, we can solve for $c_1$ by rearranging:

\begin{equation}
\left|\mathbf{\beta}-c_1\mathbf{\hat{c}}\right| = \left|\mathbf{i}\right| = 1,
\end{equation}

\noindent Which then allows the evaluation of $\mathbf{i}$.  We can then make a similar calculation for $c_2$ and $\mathbf{f}$:

\begin{equation}
\left|\mathbf{i}+c_2\mathbf{\hat{c}}\right| = \left|\mathbf{f}\right| = 1,
\end{equation}

\noindent And hence obtain $\mathbf{f}$, $c_2$ and the chord length $c$.  With the chord length determined, this gives the central angle $\theta$ and ultimately $\Sigma_s$.

\subsection{Breaking the Degeneracies}

\noindent To recapitulate, the Shkadov transit has 3 measurements which are degenerate.  The first is the transit ingress timescale $\tau_{ing}$, which is degenerate in $r_p$ and $b=\beta_2$; the second is the effective transit depth $\tilde{\delta}$, which is degenerate in $r_p$ and $\Sigma_s$; the third is the Shkadov egress timescale $\tilde{\tau}_{eg}$, which is degenerate in $r_p$ and $\xi$.  Consequently, all three measurements are interdependent - measuring one constrains the values the others may take.  

Assuming a value for $\xi$ allows the observer to use $\tilde{\tau}_{eg}$ to obtain $r_p$, constraining $b=\beta_2$ from $\tau_{ing}$ and independent measurements of $R_*$.  Measuring $\beta_1$ from the midpoint of Shkadov egress allows the calculation of $\Sigma_s$, which must be consistent with the transit depth $\tilde{\delta}$.  

Therefore, model fitting of three measurements simultaneously can break the degeneracy.  Bayesian methods are used commonly in parameter fitting of standard transit curves (e.g. \cite{Veras2009, Gazak2012, Kipping2011, Winn2011} and references within), so current techniques can be modified to accommodate the fitting of the Shkadov transit.

\subsection{Masking the Shkadov Signal}

\begin{figure}
\begin{center}
\includegraphics[scale = 0.4]{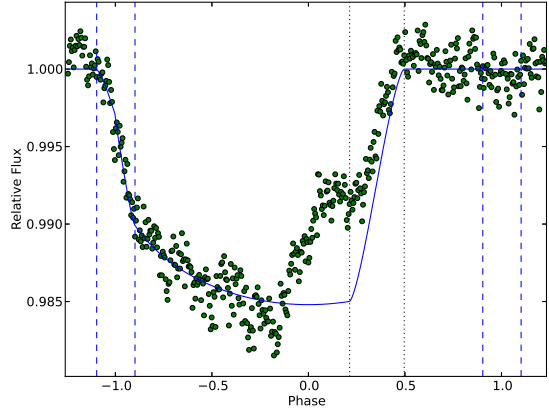} 
\caption{The effect of noise on detecting Shkadov transits.  The curve shows the analytically determined Shkadov transit curve, and the dashed and dotted lines refer to the ingress and egress points for normal and Shkadov transits, as in previous figures.  The points represent the analytic curve with added $1/f$ noise and a starspot placed near the beginning of Shkadov egress. \label{fig:noisy_shkadov}} 
\end{center}
\end{figure}

\noindent Until now, we have considered idealised transit curves that are free of the sources of noise and error that real transit curves possess.  This can include (but is not limited to) the underlying Poisson noise produced by photon processes, cosmic ray hits, scintillation effects produced by atmospheric turbulence, or other instrumental or calibration effects that (in combination) produce a background of time-correlated noise to all transit curves.  Perhaps more dangerously, the presence of starspots can produce ``dents'' in the transit light curve, as the planet covers the starspot, which is by definition a cooler portion of the stellar disc  (see e.g. \cite{Kipping2012a, Tregloan-Reed2012} for efforts to model this phenomenon).  When an observer is trying to observe the the duration of Shkadov egress, the presence of a starspot can make this extremely difficult.  Figure\ref{fig:noisy_shkadov} illustrates this problem.  We have added time-correlated $1/f$ noise to the theoretical curve, produced by generating Gaussian white noise and applying the ``pinkening'' filter of \cite{Kasdin1995}.  To complicate matters further, we add a ``starspot'' to the curve, modelled by simply adding a Gaussian with a peak value slightly less than that of the planet to the curve.  While not a particularly accurate starspot model, it demonstrates that unfortunately positioned noise sources can make the extraction of necessary Shkadov parameters particularly difficult.

\section{Discussion \label{sec:discussion}}

\subsection{Probability of Detecting a Shkadov Transit}

\noindent Let the following events be defined:

\begin{itemize}
\item T - the system possesses a detectable transiting planet with a measurable curve
\item S - the system possesses a Shkadov thruster
\item D - the thruster is oriented such that its presence can be detected in the transit curve.
\end{itemize}

Given $T$ and $S$, we can calculate the probability that the thruster will be detectable in the transit curve, $P(D|T,S)$.  The thruster is a spherical segment of azimuthal angle $2 \psi$, and presumably of polar angle $\pi$ (as this will maximise its thrust for any given $\psi$). 

\begin{figure}
\begin{center}
\includegraphics[scale = 0.3]{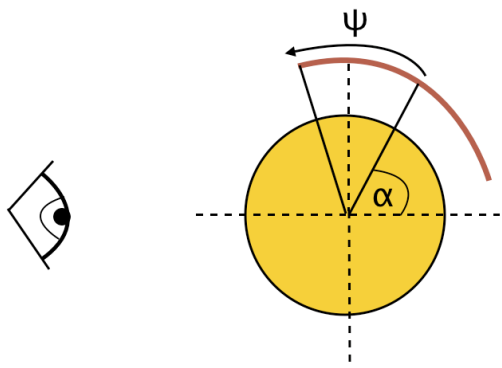} 
\caption{Orientation of the thruster to the line of sight, as defined by the angle $\alpha$. \label{fig:prob_shkadov}} 
\end{center}
\end{figure}

\noindent For the observer to see a thruster (with $\xi=0^{\circ}$), then the position angle of the midpoint of the thruster $\alpha$, defined along the line of sight of the observer (Figure \ref{fig:prob_shkadov}) must be in the range $[\pi/2 - \psi, 3\pi/2 + \psi]$.  Therefore
\begin{equation}
P(D|T,S) = P( \pi/2 - \psi  < \alpha < 3\pi/2 +\psi) = \int^{3\pi/2 + \psi}_{\pi/2 - \psi} d\alpha',
\end{equation}

\noindent which gives

\begin{equation}
P(D|T,S) = \frac{\pi + 2\psi}{2\pi}.
\end{equation}

\noindent The maximum thrust achievable requires $\psi=\pi/2$, which would give a probability of effectively 1.  In this case, there is only one possible configuration which does not give a Shkadov transit, where the mirror is positioned precisely behind the stellar disc, or equivalently $\alpha=0$ exactly (although this particular configuration still allows for occultations of the planet by the Shkadov thruster - see following section).  If $\psi=0$, then the probability is 0.5 - the thruster is now an infinitely thin line across the stellar disc, which is either on the observer-facing side of the star or the other.  In their calculations of thruster-assisted stellar motion, \cite{Badescu2000} use $\psi = 30^{\circ} = 0.523$ rad, giving a probability $P(D|T,S) = 2/3$.  In the case where $\xi \neq 0^{\circ}$,  we may expect that this probability will increase, as the total azimuthal extent of the thruster will be larger.  

Of course, this conditional probability is less relevant than $P(D)$, which is given by

\begin{equation}
P(D) = P(D|T,S) P(S|T) P(T).
\end{equation}

\noindent The probability that a system possesses a Shkadov thruster, given that the system possesses a transiting exoplanet, $P(S|T)$, is not well constrained at all.  We do not know to sufficient confidence how many planetary systems possess intelligent life, and even less clear is how many of these intelligent beings would wish to build a Shkadov thruster. $P(T)$ is a combination of the probability that the star system has any planets at all, and the probability that the plane of a planet's orbit aligns sufficiently well with the observer's line of sight, which for a circular orbit is 

\begin{equation}
P_{align} =  \frac{R_* + r_p}{a}
\end{equation}

\noindent We can hazard a guess at $P(D)$ by appealing to parameter estimates for the Drake equation to obtain $P(S|T)$, assuming all stars have planets and $P_{align}$ is of $\sim$ 0.005 for an Earth orbiting a Sun-like star at 1 AU.  If a fraction $f_i$ of planetary systems possess intelligent life, a fraction $f_c$ of intelligent species choose to communicate, and a fraction $f_T$ of intelligent species decide to build Shkadov thrusters, then we can make an educated guess of 

\begin{equation}
P(S) = P(S|T) P(T) \approx 10^{-3} f_i f_T.
\end{equation}

\noindent Given that recent surveys suggest that $f_i f_c <0.01$ in the solar neighbourhood \cite{Siemion2013}, and $f_T$ is unconstrained, even for the very unlikely case of $f_T=1$, the probability of detection is likely to remain around $10^{-4} - 10^{-5}$.  More realistic values of $f_T < 0.01$ suggest that the probability of detection has an upper limit of $<10^{-6}$.  There are $\sim 10^3$ exoplanets detected to date, and only a fraction of the planet population in the vicinity of the Earth will be amenable to detection and characterisation via exoplanet transits, which suggests that the number of locally detectable Shkadov thrusters will be small.  This reasoning is sufficient to rule out funding dedicated searches for Shkadov transits, but exoplanet transit surveys are likely to continue for the foreseeable future (see next section), and SETI scientists will have many opportunities to study public catalogues of data at low cost to attempt the detection of Class A stellar engines.

\subsection{Limitations of the Analysis}

\noindent In this work, we have made several simplifying assumptions for the sake of more clearly illustrating the Shkadov transit as a potential detection method.  In doing so, we have ignored aspects of the Shkadov transit curve that may prove important in future efforts to characterise it.  

Perhaps most glaringly, we have ignored thermal emission from the planet in this analysis.  By doing so, we discount the possibility of detecting secondary eclipses or occultations, as the planet passes behind the star.  Depending on the orientation of the thruster, a secondary eclipse may be artificially lengthened if the thruster obscures the planet as it moves into occultation, or the mirror may reflect the night-side thermal emission back toward the observer, leading to a secondary eclipse curve that is ``unbalanced'': the thermal emission at the thruster side will appear enhanced compared to the non-thruster side.  This may prove to be another means by which a Shkadov transit can be characterised.

We have assumed circular orbits in this analysis.  The eccentricity of the planet's orbit plays an important role in characterising the transit curve.  In particular, it can produce inequalities in the ingress/egress timescale of the transit.  While in practice the difference amounts to less than a few percent \cite{Winn2011,Kane2012a}, highly eccentric orbits will confuse with the Shkadov signal.  The extent to which this confusion will occur depends both on the eccentricity and the orbital longitude of periastron.  Indeed, it may be the case that to fully characterise the Shkadov transit in this case, the secondary eclipse will be required to measure the difference between the impact parameter of the transit, $b_{tra}$, and the equivalent impact parameter for the occultation, $b_{occ}$, to constrain $e$.

Amongst our assumptions, we have required that the stellar radius $R_*$ remains measurable despite the presence of the thruster.  Stellar atmosphere modelling given external information on the star's properties remains the primary means of inferring $R_*$, but these models are constructed in the absence of technological interference with the stellar environs.  The engine's presence artificially increases the effective temperature of the star.  As the effective temperature is a key input variable in most stellar atmosphere models, there is a possibility that the radius will be misestimated.  It is not immediately clear what radius the star would relax to after the thruster is placed, and it is possible that the star will present an increased oblateness depending on its rotation.  In any case, using simple stellar atmosphere models may be insufficient (or at least inaccurate).

What is to be done instead? Interferometric measurements \cite{Baines2007} or asteroseismology \cite{Stello2009} may prove to be of assistance.  Indeed, these measurements may be sufficient to confirm the existence of the Shkadov thruster through its effects on the star's structure.  The unusual transit curve would merely serve as a flag to SETI scientists to require follow-up with one of these techniques! 

It may even be the case that stars that do not possess transiting planets may still possess a Shkadov transit-esque signal, depending on the strength of their starspots.  Indeed, the screening of one side of the stellar disc may set up a permanent Rossiter-McLaughlin effect, which would be detectable in the net spectral line profile of the star (cf. \cite{Ohta2005,Winn2005}).  While perhaps weaker than the signals produced in systems with exoplanet transits, future efforts to detect Shkadov thrusters, or projects which carry out photometric monitoring of stars, should be cognisant of this potential signal.  


\subsection{Detectability in Current and Future Surveys}

\noindent A crucial factor in the ability to characterise Shkadov transits is the need for radial velocity follow-up, otherwise the semi-major axis of the planet's orbit will remain undetermined.  Without this piece of information, the unmodified ingress timescale $\tau_{ing}$ cannot be used to produce the degenerate $k/\sqrt{1-b^2}$ term, and the full deconstruction of the Shkadov transit is not possible.

If we are to identify current and future surveys that are likely to be able to yield fully characterised Shkadov transits, we should therefore focus on surveys that are amenable to radial velocity follow-up.  The formidable success of the \emph{Kepler} space telescope in detecting a large number of transiting exoplanet candidates \cite{Batalha2013} might lead us to think of it first, and indeed instruments such as HARPS-N have been designed with radial velocity follow-up of Kepler targets as a primary objective \cite{Cosentino2012,Latham2013}.  However, parts of the Kepler sample will be too faint for current ground-based instrumentation to perform radial velocity follow-up \cite{Wheatley2013}.

Future surveys with a wide, shallow observing strategy such as TESS \cite{Ricker2010}, or PLATO \cite{Catala2011} are designed with radial velocity follow-up in mind.  Selecting bright stellar candidates will allow for better ground based measurements of the radial velocity, and provide a much larger sample of candidates to search for Shkadov transits.  At the most extreme end, the GAIA satellite will provide a very large sample of low cadence measurements of stellar photometric and spectroscopic variations, providing a catalogue for follow-up missions to produce transit candidates \cite{Dennefeld2011}.

The detection of a Shkadov transit is not contingent on high precision or sensitivity beyond being visible to RV instruments - although, depending on the Shkadov egress timescale, which is typically more abrupt than a standard transit egress, cadence may be a factor.  If a candidate Shkadov transit is detected, missions designed to accurately characterise exoplanet transits to probe atmospheres, such as CHEOPS \cite{Broeg2013} or EChO \cite{Tinetti2012} could be used to confirm that the curve does indeed show a Shkadov transit. In particular, these instruments can investigate the wavelength dependence of the Shkadov component. The Shkadov egress point may slightly shift with increasing wavelength due to diffraction, but we may expect the wavelength dependence of the unobscured transit ingress to be stronger.

Despite this, if the aim is merely to detect the Shkadov transit, and not to characterise its properties, then virtually any planetary system where a transit can be detected can yield evidence of a Shkadov transit.  In this sense, searching for Shkadov transits can act as a means of flagging planetary systems for further SETI studies using other techniques.

\section{Conclusions \label{sec:conclusion}}

\noindent We have outlined a means by which Class A stellar engines (or Shkadov thrusters) can be detected in exoplanet transit curves, if the stellar engine partially obscures the stellar disc during the transit.  This constitutes a new serendipitous SETI detection method, which does not require the civilisation to be intentionally transmitting, and is present in the optical/infrared over a broad wavelength range.  Also, as transit curves are periodic in nature, the signal is by definition repeatable.  

We have shown the shape of the light curve in the presence of a Class A stellar engine, and demonstrated the deconstruction of the curve into the properties of both the transiting planet and the thruster (provided radial velocity follow up studies of the system are possible and the star's radius remains estimable).  We note that both the primary and secondary eclipses may be affected, and that interferometric and asteroseismology measurements may help to characterise the nature of the stellar engine.

However, we also note that the detection of the features in the light curve produced by a Shkadov transit can be easily masked by phenomena such as starspots.  As such, SETI scientists attempting to characterise Shkadov transits will require an excellent grasp of the sources of noise present in the light curve.  Other measurements of stellar properties from radial velocity studies, interferometric imaging and asteroseismology may prove crucial in the full deconstruction of the properties of the stellar engine.

Exoplanet transit missions are of paramount importance in understanding the physical processes of planet formation, and are likely to remain so in the future.  As such, while the probability of detecting a stellar engine is likely to be very low, current and future instruments will provide a large dataset from which to mine.  We believe this will prove to be an extremely useful tool in constraining the population of Class A stellar engines in the Milky Way.

\section*{Acknowledgments}

\noindent The author gratefully acknowledges support from STFC grant ST/J001422/1, and thanks Caleb Scharf for useful comments.

\bibliographystyle{ieeetr} 
\bibliography{shkadov_transit}

\end{document}